# Insight into the interface between $Fe_3O_4$ (001) surface and water overlayers through multiscale molecular dynamics simulations

Hongsheng Liu[1,2], Enrico Bianchetti[1], Paulo Siani[1], Cristiana Di Valentin[1*]

1. Dipartimento di Scienza dei Materiali, Università di Milano-Bicocca

via Cozzi 55, 20125 Milano Italy.

2. Key Laboratory of Materials Modification by Laser, Ion and Electron Beams, Dalian University of Technology, Ministry of Education, Dalian 116024, China

## Abstract

In this work we investigate the $Fe_3O_4$ (001) surface/water interface by combining several theoretical approaches, ranging from a hybrid functional method (HSE06) to density-functional tight-binding (DFTB) to molecular mechanics (MM). First, we assess the accuracy of the DFTB method to reproduce correctly HSE06 results on structural details and energetics and available experimental data for the adsorption of isolated water, dimers, trimers, etc. up to a water monolayer. Secondly, we build two possible configurations of a second and a third overlayer and perform molecular dynamics simulations with DFTB, monitoring the water orientation, the H-bond network, and ordered water structures formation. To make our models more realistic, we then build a 12-nm thick water multilayer on top of the $Fe_3O_4$ (001) surface slab model, which we investigate through MM molecular dynamics. The water layers structuring, revealed by the analysis of the atomic positions from a long MM-MD run for this large MM model, extends up to about 6-7 Å and nicely compares with that observed for a water trilayer model. However, MM and DFTB MD simulations show some discrepancy due to the poor description of the Fe---$OH_2$ distance in MM that calls for further work in the parametrization of the model.

---

[*] Corresponding author: cristiana.divalentin@unimib.it

1. **Introduction**

Interfacial solid surface/water effects and interactions are crucial for many fundamental and technological processes. A molecular-level understanding of water adsorption, dissociation, and clustering on model surfaces of metal oxides is currently achievable by combining experimental techniques, such as scanning tunneling microscopies, photoemission spectroscopy and temperature-programmed desorption, with quantum chemical calculations. However, identifying the water structuring in overlying layers is still a challenge for experiments, whereas it could be addressed by molecular dynamics (MD) simulations.

The water adsorption on the $Fe_3O_4$ (001) surface, one of the most important low-index facets, has been previously studied through various experimental approaches.[1,2,3,4,5,6,7,8] The temperature programmed desorption analysis detected three water desorption peaks at 320, 280 and 225 K that were assigned to chemisorbed water molecules.[1] At low water vapor pressure and at room temperature, chemisorbed water can only dissociate on defects sites, whereas at increasing partial pressure an increasing number of OH species were identified by X-ray photoemission (XPS).[2,3] Dissociated water was also observed by scanning tunneling microscopy (STM).[4] At high water coverage, more experimental indications exist that a mixed dissociated/undissociated adsorption mode is established.[3,5,6]

In order to simulate the water adsorption on the $Fe_3O_4$ (001) surface, a correct surface model must be available. Only in 2014, Parkinson and co. proposed a new reconstructed surface model[9] that agrees well with the surface X-ray diffraction data,[10] with the low energy electron diffraction (LEED) pattern[9] and could explain the site preference of deposited Au adatoms on the surface.[9] The stacking sequence in the [001] direction is of A layers containing tetrahedral Fe ions ($Fe_{tet}$) and B layers containing O and octahedral Fe ions ($Fe_{oct}$). The reconstructed surface model (labeled as SCV) presents a B layer terminated $Fe_3O_4$ (001) surface with an extra interstitial $Fe_{tet}$ atom in the second layer and two $Fe_{oct}$ vacancies in the third layer per ($\sqrt{2}\times\sqrt{2}$)R45° unit cell. Since SCV was discovered, two computational works have concomitantly appeared in the literature where water adsorption on such reconstructed surface was modeled by density functional theory calculations. In one of these studies by some of us,[11] adsorption of one to four water molecules on both the bulk-terminated and the SCV $Fe_3O_4$ (001) surface unit cell models has been investigated by hybrid density functional calculations (HSE06).[12] We have shown that in certain experimental conditions of water partial pressure and temperature, the hydrated bulk-terminated surface may become more stable than the SCV one, that, however, is the most stable in a wide range of experimental conditions. The other study is a combined experimental and computational work,[13] based on Hubbard-corrected DFT (DFT+U) calculations. In this work, it was shown that partially dissociated water dimers and trimers remain isolated because of the large distance between surface $Fe_{Oct}$ along a row and between rows. As the coverage increases to one monolayer, i.e. up to eight water molecules per unit cell of the reconstructed SCV surface model, ring-like H-bonded networks were observed.

The next step along the investigation of the magnetite surface interface with water is to go beyond the first chemisorbed water monolayer and to consider further water overlying layers, which is the main aim of this work. It is not necessarily valid that the most stable configuration for one monolayer of water on the surface is maintained when an overlying water layer forms, especially if the interaction energy of the water molecules in the first layer is competitive with the energy of the interactions that would be established with the water molecules of the second layer. Here, we investigate by means of molecular dynamics (MD), based on the self-consistent charge density-functional tight-binding (SCC-DFTB) method,[14] the temperature effect on the arrangement of an increasing number of water molecules on the $Fe_3O_4$ (001) surface. First, we assess the accuracy of the SCC-DFTB method for the water adsorption on magnetite surface through the comparative analysis of the adsorption energy and structure of one, two, three etc. water molecules up to one monolayer with respect to the corresponding data from a hybrid density functional method (HSE06). Then, we run MD simulations of time length of about 50 ps at 300 K for the bilayer and trilayer of water molecules on the surface, in order to monitor water orientation during the dynamics, formation of H-bonding networks, and eventual appearance of ordered water structures. On top of that, we also perform MD simulations based on molecular mechanics, which, given the reduced computational cost, allow us to improve the interface model from a water trilayer to a thick water multilayer (12 nm), i.e. bulk water, on the $Fe_3O_4$ (001) surface. This comparative multiscale approach involving *ab initio*, tight binding and molecular mechanics methods allows not only spanning different space and time scales, but also to investigate with the proper accuracy both chemical and physical phenomena.

2. **Computational Details**

Most of the calculations are performed using SCC-DFTB method implemented in DFTB+ package.[15] The SCC-DFTB is an approximated DFT-based method that derives from the second-order expansion of the Kohn-Sham total energy in DFT with respect to the electron density fluctuations.[14] The SCC-DFTB total energy can be defined as:

$$E_{tot} = \sum_i^{occ} \varepsilon_i + \frac{1}{2}\sum_{\alpha,\beta}^{N} \gamma_{\alpha\beta}\Delta q_\alpha \Delta q_\beta + E_{rep} \quad (1)$$

where the first term is the sum of the one-electron energies $\varepsilon_i$ coming from the diagonalization of an approximated Hamiltonian matrix, $\Delta q_\alpha$ and $\Delta q_\beta$ are the induced charges on the atoms α and β, respectively, and $\gamma_{\alpha\beta}$ is a Coulombic-like interaction potential. $E_{rep}$ is a short-range pairwise repulsive potential. More details about the SCC-DFTB method can be found in Refs. 16, 17 and 18. DFTB will be used as a shorthand for SCC-DFTB.

For the Fe-Fe and Fe-H interactions, we used the "trans3d-0-1" set of parameters, as reported previously.[19] For the O-O, H-O and H-H interactions we used the "mio-1-1" set of parameters.[14] For the Fe-O interactions, we used the Slater-Koster files fitted by us previously,[20] which can well reproduce the results for magnetite bulk and surfaces from HSE06 and PBE+U calculations.[21] To properly deal with the strong correlation effects among Fe 3d electrons,[22] DFTB+U with an effective U-J value of 3.5 eV was adopted according to our previous work on magnetite bulk and (001) surface.[11,23] The convergence criterion of $10^{-4}$ a.u. for force was used during geometry optimization, and the convergence threshold on the self-consistent charge (SCC) procedure was set to be $10^{-5}$ a.u. The k points generated by the Monkhorst–Pack scheme were chosen to be 6×6×1.

DFTB+U molecular dynamics were performed within the canonical ensemble (NVT) with a time step of 1 fs. An Andersen thermostat[24] was used to target the desired temperatures. To simulate the temperature annealing processes, the system was quickly heated up to 400 K (within 1 ps) and then kept at 400 K for 4 ps, and then cooled down slowly to 50 K when no further structural changes were expected. The total simulation time is 30 ps. Furthermore, DFTB+U molecular dynamics at constant temperature (300 K) was performed to investigate the effect of temperature on the water molecules orientation. The total simulation time is 50 ps. For all the molecular dynamics simulations, the k points generated by the Monkhorst–Pack scheme were chosen to be 4×4×1.

To well describe the hydrogen bonds, a modified hydrogen bonding damping (HBD) function was introduced with a $\zeta = 4$ parameter[25] for the DFTB+U calculations of water monolayer, bilayer, and trilayer. We further checked that the structure for the adsorption of four water molecules is not affected by the inclusion of the van der Waals correction (DFTB+D3)[26,27]. Since the variations are within 0.1 Å, no correction will be presented in the following.

To assess the reliability of DFTB+U results, hybrid functional calculations (HSE06) were also carried out using the CRYSTAL17 package[28,29] at the DFT level of theory for a comparative analysis. In these calculations, the Kohn−Sham orbitals are expanded in Gaussian-type orbitals (the all-electron basis sets are H|511G(p1), O|8411G(d1) and Fe|86411G(d41), according to the scheme previously used for $Fe_3O_4$).[11,23] The convergence criterion of 0.023 eV/Å for force was used during geometry optimization and the convergence criterion for total energy was set at $10^{-6}$ Hartree for all the calculations. The k points generated by the Monkhorst–Pack scheme were chosen to be 3×3×1 since total energy difference was found to be below 1 meV when compared with larger grids up to 6×6×1. According to a previous report,[30] the inclusion of the van der Waals correction (DFT+D2)[31] only slightly changes the adsorption energy of water on the $Fe_3O_4$(110) surface, so no van der Waals correction is included in this work. The basis set superposition error (BSSE) was evaluated in one sample case by using the counterpoise correction method (see Table 1).[32]

For the $Fe_3O_4$(001) surface, the SCV model is considered. According to previous reports, this structural model is more stable than other models.[9,11] We used the same structure presented in our previous works[11,20,23] that is a (1×1) 17-layer slab with inversion symmetry. In the z direction a vacuum of more than 12 Å was introduced to avoid the spurious interaction between the periodic sides of the slabs. Five layers in the middle of the slab are kept fixed to the bulk positions, whereas the other layers are fully relaxed. For water adsorption, molecules were put on both sides of the slab.

To evaluate the stability of water adsorption on the $Fe_3O_4$(001) surface, the adsorption energy per water molecule ($E_{ads}$) was calculated as follows:

$$E_{ads} = (E_{total} - E_{slab} - N_{H_2O} \times E_{H_2O})/N_{H_2O} \quad (2)$$

where $E_{total}$ is the total energy of the whole system (surface slab and adsorbed water), $E_{slab}$ is the energy of the $Fe_3O_4$(001) surface slab, $N_{H_2O}$ is the number of water molecules adsorbed and $E_{H_2O}$ is the energy of one isolated water molecule. This formula provides a value for the adsorption energy which is averaged on all the water molecule.

To evaluate the cohesion between different water layers, the adhesion energy per water molecule ($E_{adh}$) was calculated as follows:

$$E_{adh}^{2layer} = (E_{total} - E_{monolayer} - 2 \times N_{H_2O}^{layer} \times E_{H_2O})/2 \times N_{H_2O}^{layer} \quad (3)$$

$$E_{adh}^{3layer} = (E_{total} - E_{doublelayer} - 2 \times N_{H_2O}^{layer} \times E_{H_2O})/2 \times N_{H_2O}^{layer} \quad (4)$$

where $E_{adh}^{2layer}$ is the $E_{adh}$ between the second water layer and the first water layer and $E_{adh}^{3layer}$ is the $E_{adh}$ between the third water layer and the second water layer. $E_{total}$ is the total energy of the whole system (surface slab and adsorbed water), $E_{monolayer}$ is the energy of the surface slab adsorbed with monolayer water, $E_{doublelayer}$ is the energy of the surface slab adsorbed with double-layer water, $N_{H_2O}^{layer}$ is the number of water molecules within one water layer (the factor 2 is for considering both sides of the slab) and $E_{H_2O}$ is the energy of one isolated water molecule.

All atomistic MM-MD simulations were carried out with the LAMMPS program (version 7 Aug 2019).[33] We made use of the CLASS2 potential style (see Ref. 34 for a full description of the COMPASS class II force field (FF). In this FF, the repulsive and dispersive van der Waals interactions are modeled by a Lennard-Jones 9-6(LJ 9-6 functional form (eq. 5), whereas the long-range electrostatic interactions are modeled by a classical Coulomb functional form (eq. 6).

$$E_{vdW} = \sum_{i,j} \epsilon_{ij} \left[ 2\left(\frac{\sigma_{ij}}{r_{ij}}\right)^9 - 3\left(\frac{\sigma_{ij}}{r_{ij}}\right)^6 \right] \quad (5)$$

$$E_{elec} = \frac{1}{4\pi\epsilon_0} \sum_{i,j} \frac{q_i q_j}{r_{ij}} \quad (6)$$

We used a slab model based on a (4×4) supercell of the $Fe_3O_4$(001) unit cell. This surface model was then hydroxylated, according to what observed in the DFTB+U calculations, through dissociation of 32 water molecules on each side of the slab. The LJ 9-6 COMPASS-FF parameters for the Fe(II), Fe(III), and O(II) atom-types were taken from Ref. 35. Partial charges for this set of atoms were derived from the aforementioned HSE06 calculations. Bonded and non-bonded parameters for the COMPASS-based three-site water model and hydroxyl group parameters were obtained in the INTERFACE force field database.[36] For different atom-types, all LJ 9-6 cross-term parameters were given by a 6$^{th}$ order combination law[37] as follows:

$$\sigma_{i,j} = \left(\frac{\sigma_i^6 + \sigma_j^6}{2}\right)^{1/6} \quad (7)$$

$$\epsilon_{i,j} = 2\sqrt{\epsilon_i \epsilon_j} \left(\frac{\sigma_i^3 \sigma_j^3}{\sigma_i^6 \sigma_j^6}\right) \quad (8)$$

Further details and tests on the MM model against DFTB+U and DFT/HSE06 data are presented in the supplementary material in the section entitled "Further computational details".

With the MM-MD method, we reinvestigated the water trilayer for comparison with the DFTB+U MD results. Then, we made use of the PACKMOL program[38] to solvate the hydroxylated slab model and set up a simulation box with 2500 undissociated water molecules on both sides of the hydroxylated $Fe_3O_4$ surface, whose repeated images are separated along the z-direction by ~12 nm-thick water multilayer. To keep the DFTB+U-optimized geometry of the hydroxylated slab model, we froze the atoms by zeroing the forces on them during all MD simulations.

To minimize the total energy of the systems, as well as avoid any atomic overlapping, we carried out a minimization phase with 500000 steps and a convergence tolerance of $10^{-8}$ for forces. An external pressure tensor was applied on the water molecules during the energy minimization to adjust the volume of the simulation box with P=1 atm. Langevin thermostat[39] heated the system to T=300 K and kept it constant during all minimization phase with an oscillation period of 0.1 ps. Equilibration phase was carried out in the isotherm-isobaric (NVT) ensemble at T=300 K with a damping parameter of 0.1 ps. Production phase explored 10ns of the conformational space under a NVT ensemble with T=300 K. Electrostatic and LJ 9-6 interactions utilized a cut-off of 10 Å, and the Newton's equations of motion were solved using the Velocity-Verlet integrator[40] with a time step of 1.0 fs.

## 3. Results and discussion
### 3.1 Assessment of accuracy of the DFTB description vs DFT/HSE06 results
#### 3.1.1 Adsorption of an isolated water molecule

When adsorbing an isolated water molecule on the (unit cell of) $Fe_3O_4$ (001) surface, we considered both undissociated and dissociated adsorption modes, where the water molecule physisorbs through the coordination of its O atom with one of the four surface 5-coordinated Fe at octahedral sites ($Fe_{oct-5c}$) ions (Fe-O distance of 2.23 Å) or chemisorbs through a heterolytic dissociation with the resulting $OH^-$ species bound to one of the four surface $Fe_{oct-5c}$ ions (Fe-O distance of 1.79 Å) and the $H^+$ bound to a nearby $O_{3c}$, respectively. All surface O atoms are three-fold coordinated, but $H^+$ is found to preferentially adsorb on those O atoms that are not directly bond to $Fe_{tet}$ atoms, which are found to be more reactive, as reported in previous works[41,42] and confirmed here. Undissociated water adsorption is preferred over the dissociated one by -0.32 eV ($E_{ads}$ = -1.14 vs -0.82 eV). The absolute value of the adsorption energy depends slightly on the method, as shown in Table 1. However, the adsorption energy difference between different configurations ($\Delta E_{ads}$) is more important and determines the interface structures. The $\Delta E_{ads}$ values obtained with different methods are reported in Table 1, which show that the DFTB+U results are in good agreement with DFT/HSE06 ones in the current work. The $\Delta E_{ads}$ values obtained with DFT/optPBE-DF+U in ref. 13 is however smaller, which may be due to the dispersion correction method used in that work.

**Table 1**. The adsorption energy of water on $Fe_3O_4(001)$ surface at different coverages (one to eight water molecules per unit cell) obtained with different methods.

| Coverage | Configurations | DFTB+U (DFTB+) | | DFT/HSE06 (CRYSTAL17)[a] | | DFT/optPBE-DF+U (VASP)[13] | |
|---|---|---|---|---|---|---|---|
| | | $E_{ads}$ (eV) | $\Delta E_{ads}$ (eV) | $E_{ads}$ (eV) | $\Delta E_{ads}$ (eV) | $E_{ads}$ (eV) | $\Delta E_{ads}$ (eV) |
| 1 $H_2O$ | Molecular | -1.14 | 0 | -0.94 | 0 | -0.64 | 0 |
| | Dissociated | -0.82 | +0.32 | -0.68 | +0.26 | -0.59 | +0.05 |
| 2 $H_2O$ | Mixed | -1.18 | 0 | -1.09 | 0 | -0.92 | 0 |
| | Molecular | -1.02 | +0.16 | -0.93 | +0.16 | -0.66 | +0.26 |
| | Dissociated | -0.73 | +0.45 | -0.50 | +0.59 | - | - |
| 3 $H_2O$ | Linear | -1.11 | 0 | - | - | -0.88 | 0 |
| | Non linear | -0.95 | +0.16 | - | - | -0.88 | 0 |
| 4 $H_2O$ | Mixed | -1.02 | 0 | -0.98 | 0 | - | - |
| | Molecular | -0.94 | +0.08 | -0.94 | +0.04 | - | - |
| | Dissociated | -0.46 | +0.56 | -0.62 | +0.36 | - | - |
| 6 $H_2O$ | | -0.84 | 0 | - | - | -0.84 | 0 |
| 8 $H_2O$ | Model I | -0.70 (-0.97)[b] | 0 | -0.97 | 0 | -0.83 | 0 |
| | Model II | -0.69 (-0.95)[b] | +0.01 | -0.90 | +0.07 | - | - |

[a] BSSE correction has been estimated for the molecular water adsorption of one water molecule to be about 0.35 eV.
[b] HBD correction included (see Computational Details).

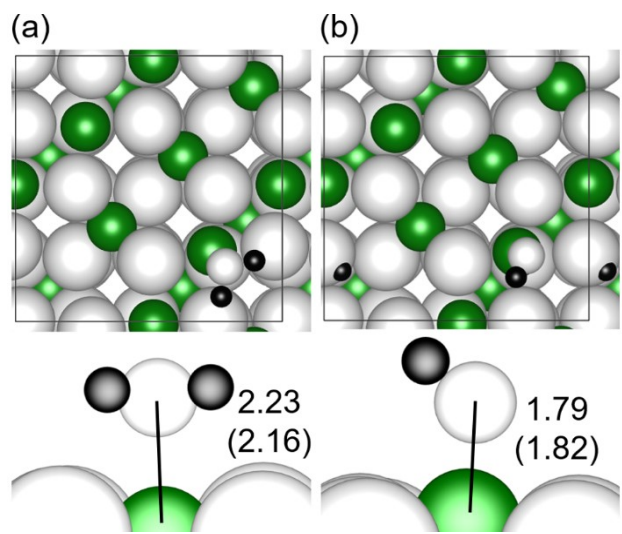

**Figure 1**. Top and side views of Fe$_3$O$_4$(001) surface with a single (a) molecular water and (b) dissociated water adsorbed on the top. The green, black, small white and big white balls represent Fe, H, O from water and O from the surface. The Fe-O$_w$ bonds are indicated by the solid lines. The bond length calculated by DFTB+U (outside the round brackets) and DFT/HSE06 (inside the round brackets) are given for each configuration. The black squares represent the ($\sqrt{2} \times \sqrt{2}$)R45° unit cell used in the calculations.

In particular, the Fe-O$_w$ (O$_w$ represents the O atom in water) distances are 2.16 and 1.82 Å with DFT/HSE06, whereas adsorption energies are -0.94 and -0.68 eV with DFT/HSE06 in this work (BSSE is evaluated to be around 0.35 eV) and -0.64 and -0.59 eV with DFT/optPBE-DF+U in a previous study for undissociated and dissociated water on Fe$_3$O$_4$ (001) surface, respectively. The Fe-O$_w$ bond length obtained with DFTB+U method also agrees well with that obtained with DFT/HSE06 method with a deviation of only about 2% to 3%.

Moreover and more importantly, all these (DFTB and DFT) results agree with the experimental observations by thermal desorption spectroscopy (TDS) and STM imaging at very low water density.[13] The experimental adsorption energy of water at low pressure is of -0.85 eV and the sparse bright spots on the surface in the STM images are assigned to OH from water.[13]

*3.1.2 Adsorption of a water dimer*

Water molecules may cluster in dimers when adsorbed on the surface. We considered four configurations, one with both undissociated molecules, two with one dissociated and one undissociated molecule (only one is shown because they are very similar) and one with both dissociated molecules (see Figure 2). The binding mode in all cases is the $O_w$ coordination to surface $Fe_{oct-5c}$ ions.

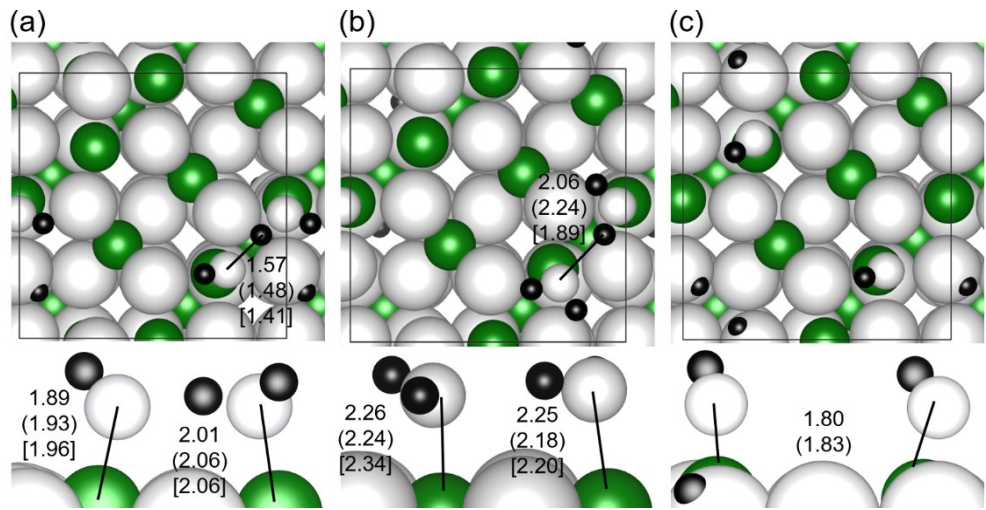

**Figure 2**. Top and side views of $Fe_3O_4(001)$ surface with two water molecules adsorbed on the top in different configurations: (a) mixed dissociated/undissociated adsorption mode, (b) molecular adsorption mode and (c) fully dissociated adsorption mode. The green, black, small white and big white balls represent Fe, H, O from water and O from the surface. The Fe-$O_w$ bonds and hydrogen bonds are indicated by the solid lines. The bond length calculated by DFTB+U (without any brackets), DFT/HSE06 (inside the round brackets) and DFT/optPBE-DF+U (inside the square brackets) are given for each configuration. The black squares represent the $(\sqrt{2} \times \sqrt{2})R45°$ unit cell used in the calculations.

As detailed in Table 1, the mixed dissociated/undissociated adsorption mode (Figure 2a) is more favorable ($E_{ads}$ = -1.18 eV) than both the molecular (Figure 2b, -1.02 eV) and the fully dissociated one (Figure 2c, -0.73 eV), in line with what computed at both DFT/HSE06 (-1.09, -0.93 and -0.50 eV, respectively) in this work and DFT/optPBE-DF+U (-0.92 and -0.66 eV, respectively) in a previous study. The $\Delta E_{ads}$ values show a high degree of consistency between different methods. This is because water is a better H-donor and OH is a better H-acceptor. Thus, their combination give rise to the strongest H-bond interaction. The H-bond distance between the dissociated (H-acceptor) and molecular (H-donor) water in Figure 2a is 1.57 Å with DFTB, which is slightly larger than what computed at DFT level of theory (1.48 and 1.41 Å with HSE06 and optPBE-DF+U, respectively). The H-bond distance between two undissociated water molecules in Figure 2b is longer, i.e. 2.06 Å with DFTB+U but 2.24 Å with DFT/HSE06 and only 1.89 Å with DFT/optPBE-DF+U.

Therefore, we conclude that through partial dissociation, water dimers become stable species on the magnetite surface characterized by a shorter Fe-OH bond and a shorter intermolecular H-bond than for molecular adsorption. The experimental STM and AFM images confirm their presence on the surface for samples that have been covered by low doses of water, as reported in ref. 13, where they appear as pairs of bright spots.

### 3.1.3 Adsorption of few water molecules (from three to six)

When a third water molecule is added to the mixed dissociated/undissociated water dimer on the $Fe_3O_4$ surface model, two possibilities have been considered that imply an interaction of the newly coming water with the dimer. In the first case, the third molecule molecularly adsorbs on a third surface $Fe_{oct-5c}$ atom in the same row along the [110] direction where the other two water molecules are adsorbed and establishes a H-bond with the OH species of the dimer, which is an excellent proton acceptor. We label this configuration as "linear" in Table 1 and Figure S1. In the second case, the third water molecule becomes involved in three H-bonds: one (H-acceptor) with the undissociated water molecule of the dimer (H-donor), one (H-acceptor) with a surface OH formed upon dissociation of one water molecule of the dimer (H-donor) and one (H-donor) with a surface O atom (H-acceptor). We label this configuration as "nonlinear" in Table 1 and Figure S1. The nonlinear configuration is less stable than the linear one by +0.16 eV, in contrast with what has been previously observed with DFT/optPBE-DF+U, where the two structures were found to be isoenergetic. Moreover, the nonlinear trimer obtained with DFT/optPBE-DF+U presents only one H-bond with the surface OH species formed upon proton transfer during water dissociation of one molecule of the dimer. The existence of linear water trimers on the surface is proved by STM and AFM images at low water partial pressure.[13]

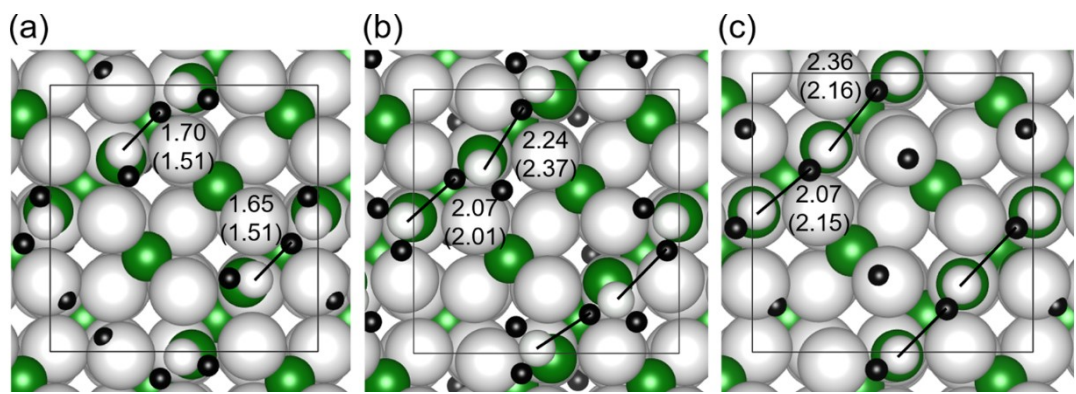

**Figure 3**. Top views of $Fe_3O_4$(001) surface with four water molecules adsorbed on the top in different configurations: (a) mixed dissociated/undissociated adsorption mode, (b) molecular adsorption mode and (c) fully dissociated adsorption mode. The green, black, small white and big white balls represent Fe, H, O from water and O from the surface. The hydrogen bonds are indicated by the solid lines. The bond length calculated by DFTB+U (without any brackets) and DFT/HSE06 (inside the round

brackets) are given for each configuration. The black squares represent the (√2 × √2)R45° unit cell used in the calculations.

When four water molecules are adsorbed on the magnetite surface, they are expected to combine in two dimers. We considered that both dimers are in the mixed dissociated/undissociated (Figure 3a), or both in the molecular one (Figure 3b) or, finally, both in the dissociated configuration (Figure 3c). As for the single dimer, the mixed adsorption mode is the most stable at both DFTB+U and DFT/HSE06 levels of theory, with similar relative energy differences. We may rationalize this with the following: water is a better H-donor than OH, and OH is a better H-acceptor than water, therefore, the mixed pair can establish the strongest H-bond interaction, as proven by the OH—O distances (1.65-1.70) reported in Figure 3. The fully dissociated configuration is rather unstable because only two of the four protons that are released can go on surface O atoms that are not directly bond to $Fe_{tet}$ atoms and are more prone to accept them. The other two protons must go on other surface O atoms that are less reactive. It is interesting to note that H-bond chains are formed along [110] direction in both structures of Figure 3b and 3c. Since both STM and AFM studies prove the presence of water dimers on the surface, it is reasonable to expect that a slight increase in water concentration will lead to an increase in the number of water dimers observed.

We also considered the possibility of having six water molecules on the surface (Figure S2) because it has been proposed in the previous work based on DFT/optPBE-DF+U calculations. The $E_{ads}$ computed in this work with DFTB+U method is in excellent agreement with the DFT/optPBE-DF+U value of -0.84 eV (see Table 1). The fifth and sixth molecules organize themselves in a way to establish a bridge, through several H-bonds, between the two dimers that lie on two different rows of $Fe_{oct-5c}$ along the [110] direction. The $E_{ads}$ of -0.84 eV is lower than what computed for the four water molecules most stable structure because the additional two molecules are not in direct contact with the surface but only establish bridging H-bonds. The atomic distances reported in Figure S2 show a good agreement between DFTB+U and DFT/optPBE-DF+U results. The existence of bridging water molecules between the hydrated $Fe_{oct-5c}$ rows along the [110] direction is confirmed by experimental AFM images.[13]

### 3.1.4 *Adsorption of one water monolayer*

The addition of two more water molecules to the model structure discussed in the previous paragraph leads to the formation of a complete first monolayer of water on the magnetite (001) surface. In this work, we are considering two different models shown in Figure 4. Meier et al. have previously proposed the model in Figure 4a (model I), whereas the model in Figure 4b (model II) is original.

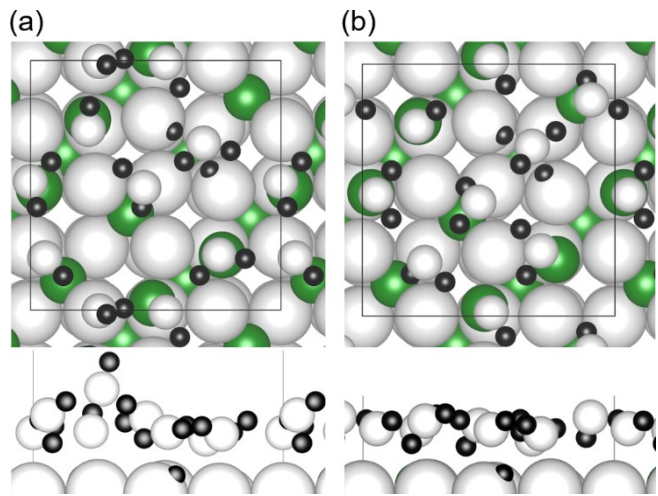

**Figure 4**. Top and side views of water monolayer adsorbed on the top $Fe_3O_4$(001) surface in two different configurations, (a) model I and (b) model II. The green, black, small white and big white balls represent Fe, H, O from water and O from the surface. The black squares represent the ($\sqrt{2} \times \sqrt{2}$)R45° unit cell used in the calculations.

We have optimized model I and then performed a simulated thermal annealing at 400 K, a subsequent cooling of the temperature and a final geometry optimization with the DFTB+U method. The resulting structure is essentially the same obtained by DFT/optPBE-DF+U calculations in ref. 13. AFM images for a water coverage of one monolayer indicate the presence of some protruding OH groups that appear as very bright isolated spots.[13]

Model II was built moving the highest water molecule in model I (see side view in Figure 4a, on the left), which is H-bonded (H-donor) to an OH species on a $Fe_{oct-5c}$ atom and does not seem to belong to the first layer, into the small hole we may observe on the right in the side view of Figure 4a. In this way, a flatter water layer is prepared. After that, we performed optimization, simulated annealing at 400 K, and cooled down for a second final optimization leading to the structure shown in Figure 4b. The reason to search for a flatter water monolayer is in the view of growing a second and then the third layer on top of it. Model II is found to be very close in energy to model I: the $E_{ads}$ is -0.69 eV to be compared to -0.70 eV computed in the case of model I. As reported in Table I, a tiny energy difference (0.01 eV) between the two models is also computed (i) when the description of the H-bonds is refined by adding the HBD correction to the DFTB+U calculations (0.02 eV) and (ii) when using DFT/HSE06 method for full atomic relaxation (0.07 eV). The DFT/HSE06 structures are very similar to the DFTB+U ones.

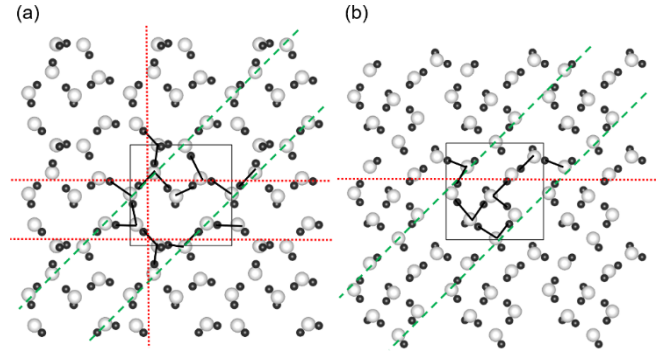

**Figure 5**. H-bond network in the monolayer water adsorbed on the $Fe_3O_4(001)$ surface for (a) model I and (b) model II, respectively. The black segments represent the H-bonds as obtained at DFTB+U level of theory. The dotted red lines indicate the directions along which the H-bond networks develop. The dashed green lines indicate the [110] directions of $Fe_{oct-5c}$ atoms on the surface where the water molecules are adsorbed. The black squares represent the $(\sqrt{2} \times \sqrt{2})R45°$ unit cell used in the calculations.

The energy difference could be attributed to the different total number of H-bonds formed, counting also those with surface atoms, which is fourteen/fourteen for model I and twelve/twelve for model II with DFTB+U and DFT/HSE06, respectively. In Figure 5, the dashed green lines indicate the [110] directions of $Fe_{oct-5c}$ atoms on the surface where the water molecules are adsorbed. The dotted red lines indicate the directions along which the H-bond networks develop. The black segments represent the H-bonds as obtained at the DFTB+U level of theory.

The overall comparison of the $Fe-O_w$ and H-bond lengths obtained with DFTB+U and DFT/HSE06 methods for the water/$Fe_3O_4(001)$ systems are listed in Table S3 and S4 in the supplementary material. We can see that the average $Fe-O_w$ bond length deviation is only 2.4% and the largest deviation is 7.2%. For the H-bond length, which is more difficult to describe well, the average deviation is 8.2%. There is one special case (four molecular water molecules on $Fe_3O_4(001)$ surface) where the DFT/HSE06 calculation does not identify a H-bond, while the DFTB+U does, resulting in a large bond length deviation of 23.8%. Considering how delicate is the H-bond network description, we judge that the DFTB+U method gives a good overall description of the water/$Fe_3O_4(001)$ interface.

To conclude section 3.1, we may point out that the comparative analysis of DFTB+U results with respect to more sophisticated methods such as DFT/HSE06 from this work and DFT/optPBE-DF+U results from a previous work[13] of another group has assessed the validity of this approximated method to describe the magnetite/water interface correctly. Therefore, in the following, we will proceed in the study of water multilayers based on this methodology, as detailed in the Computational details (i.e., using the HBD correction for an improved description of water/water interactions through H-bonds).

### *3.2 Molecular dynamics simulation of water bilayers*

We have built two models of water bilayers on the $Fe_3O_4$ (001) surface (see Figure 6) starting from the two models of water monolayers considered in the previous section, by adding other eight water molecules resulting in a bilayer of sixteen water molecules. In Figure 6, the molecules assigned to the first and the second layer, according to the construction approach just described, are shown in different colors in the space-filling model. We present the structures obtained 1) through direct full atomic relaxation and 2) through consecutive molecular dynamics (T was increased up to 400 K and then the system was cooled down again) and full atomic relaxation. First, we can observe that the second layer on model I presents a protruding OH bond from one water molecule, whereas in model II all water molecules lie on the horizontal plane, with some OH bonds pointing downwards towards the first layer. These observations hold for the structures obtained before (a and b) and after (c and d) the MD simulation. After MD simulation at 400 K and cooling for an overall time of 30 ps of model I, one of the water molecules of the second layer (on the right) moved to the first layer, whereas one of the first layer goes in the second layer (on the left). In model II, the rearrangements upon thermal treatment of the sample are smoother, although one can observe in Figure 6d that a water molecule of the second layer rotates to better fit with the molecules of the first layer.

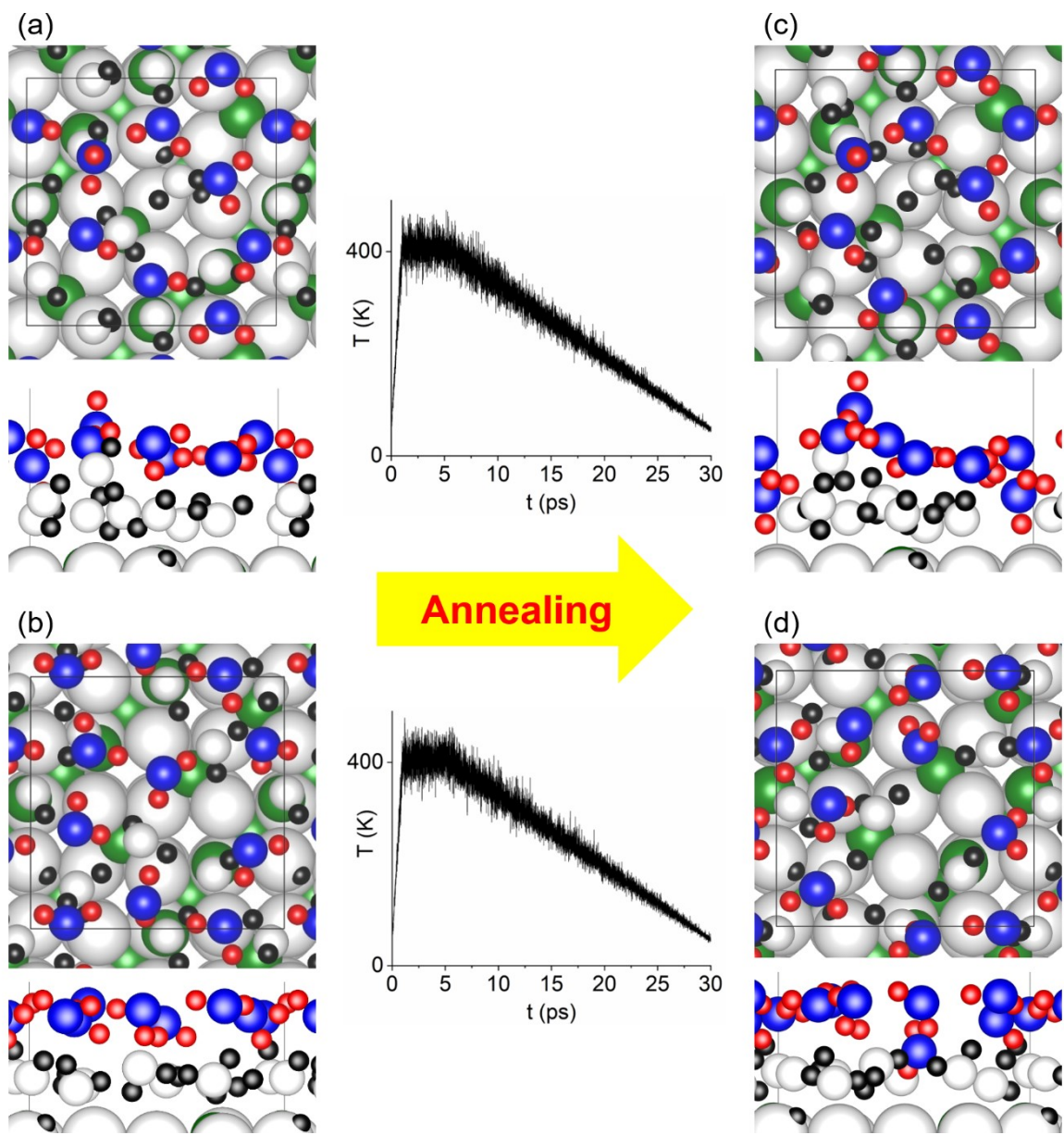

**Figure 6**. Top and side views of water bilayer adsorbed on the top $Fe_3O_4(001)$ surface in two different configurations. (a) and (c) are model I before and after annealing, respectively. (b) and (d) are model II before and after annealing, respectively. The green, black, red, big white, small white and blue balls represent Fe, H from the first water layer, H from the second water layer, O from the surface, O from the first water layer and O from the second water layer. The black squares represent the $(\sqrt{2} \times \sqrt{2})R45°$ unit cell used in the calculations. Simulated annealing temperature profiles are inserted in the middle.

In order to get insight into the structure of the water layers, we have analyzed the water molecules distance and orientation from the surface by plotting the distribution functions 1) of the z coordinate of the water O atoms (Figure 7a and 7c) and 2) of the angle between the OH bonds direction and the surface normal (Figure 7b and 7d), considering all the structures obtained at every step of the production run in the DFTB+U molecular dynamics simulation at 300 K for 50 ps. In Figure 7a and c, on the left we present the distribution of the z coordinates of the O atoms and on the right we present their time dependence for all the sixteen water molecules. The graphs indicate clearly that the second layer is for both the bilayer models less compact than the first (average z-value for O atoms in the second layer of model I [4.39±0.62 Å] and of model II [4.27±0.61 Å] vs that in the first layer of model I [2.27±0.33 Å] and of model II [2.09±0.22 Å], respectively. Different layers here are defined according to the peaks in the distribution of the z coordinates of the O atoms as shown in Figure 7a and c.). In model I, there is one molecule on the top of the second layer that results to be quite detached (see red arrow). We also observe a sort of empty corridor between the first and the second layer. In Figure 7b and 7d, we may observe the orientation of the molecules in the two layers (grey and blue dots) for the two models proposed. In the first layer of both models, the water molecules are oriented in the layer plane but they also point downwards toward the surface atoms and upwards toward the second layer. This is an indication that they are forming all types of H-bonds: intralayer, interlayer and with surface O atoms. In the second layer of both models, the situation is clearly different, due to the presence of vacuum. The molecules clearly do not point upwards but are mostly lying in the layer plane or some point downwards toward the water molecules of the first layer. However, there is one OH from one water molecule in the second layer of model I that always points upwards during the MD simulation, which results in a peak at $\cos\theta = 1$ in the angle distribution plotting for the second water layer (Figure 7b).

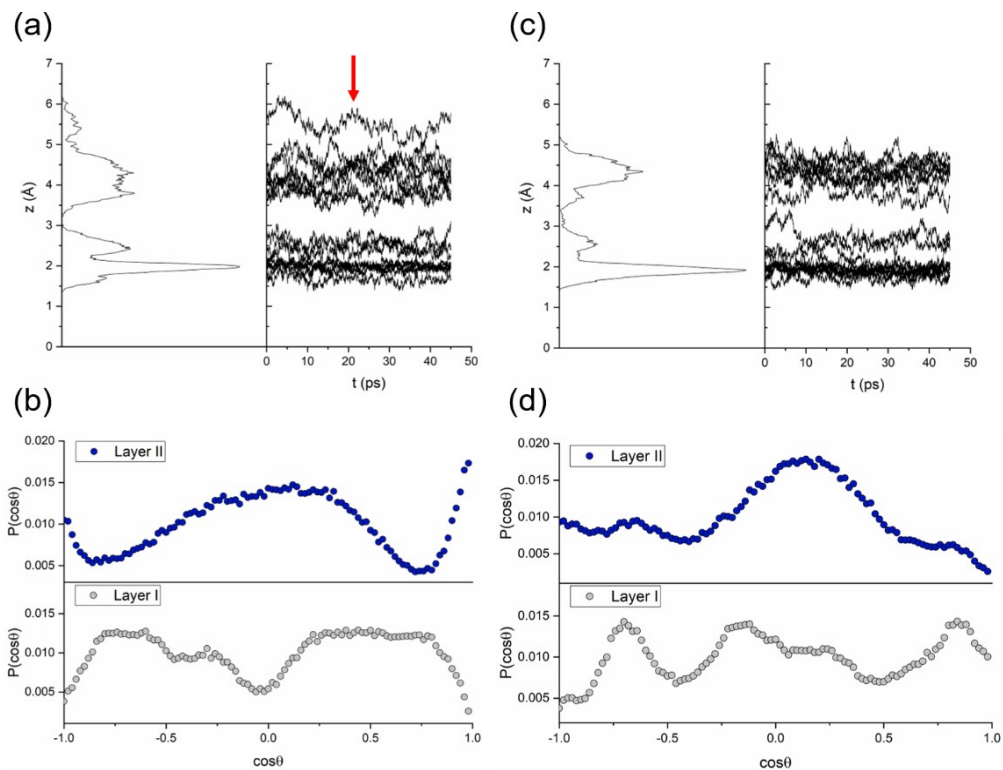

**Figure 7**. (a) and (c) are the distribution and time evolution of z (perpendicular distance from the surface) for the water oxygens of an adsorbed water bilayer on the $Fe_3O_4$(001) surface for model I and model II, respectively. z is referred to the average value of z of the four $Fe_{Oct-5c}$ atoms on the surface. (b) and (d) are probability distribution P(cosθ) of the angle θ between the O-H bond vector and the normal of the surface for the molecules of the water bilayer adsorbed on the $Fe_3O_4$(001) surface for model I and model II, respectively. Cosθ equals to 1 means that the O-H bond is directed upwards, whereas a value close to -1 means that the O-H bond is directed downwards.

The adsorption energy per water molecule ($E_{ads}$) is -0.78 eV for both models I and II of bilayer structures. Therefore, the energy difference of 0.02 eV in favor of model I at the one monolayer coverage is fully recovered when a second water layer is put on top. For this reason, the adhesion energy per molecule ($E_{adh}$) of the second layer is slightly larger for model II than for model I (-0.60 eV vs. -0.59 eV). These adhesion energy values per molecule are lower than those computed for the first layer in the monolayer structure because the latter is directly bound to the magnetite surface, whereas the second layer is bound to a first water layer, which is clearly a weaker interaction.

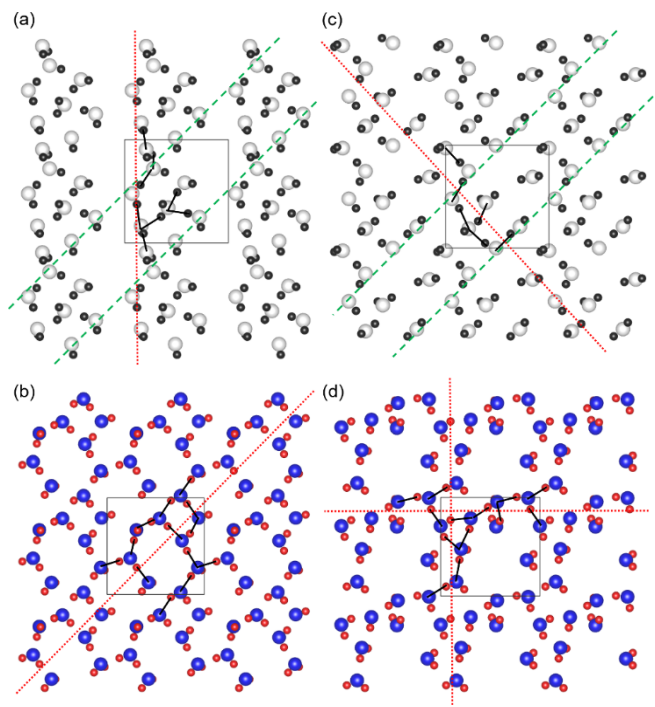

**Figure 8**. H-bond network in the first ((a) and (c)) and second ((b) and (d)) water layer adsorbed on the $Fe_3O_4$(001) surface for model I (left) and model II (right). The black segments represent the H-bonds as obtained at DFTB+U level of theory. The dotted red lines indicate the directions along which the H-bond networks develop. The dashed green lines indicate the [110] directions of $Fe_{oct-5c}$ atoms

on the surface where the water molecules are adsorbed. The black squares represent the $(\sqrt{2} \times \sqrt{2})R45°$ unit cell used in the calculations.

There is a simple reason why the two models are isoenergetic: the two bilayer structures present the same number of H-bonds since we observe seven and six H-bonds in the first water layer, ten and nine H-bonds in the second layer, eight and ten H-bonds between the two layers, and three with the surface O atoms, for model I and model II respectively. The sum of all these H-bonds is twenty-eight for both the bilayer configurations.

We have analyzed whether there is some H-bond network existing in these two model structures, as presented in Figure 8. We investigated each layer, separately and clearly observe in-plane H-bond networks that become infinite through the repetition of periodic unit cells. The interlayer H-bonds serve only to establish some layer adhesion, as discussed above.

### *3.3 Molecular dynamics simulations of water trilayers*

Two models of water trilayers on the $Fe_3O_4$ (001) surface are built starting from the two models of water bilayers obtained (after molecular dynamics and full atomic relaxation) in the previous section, by adding further eight water molecules, which results into a trilayer of twenty-four water molecules. In Figure 9, the molecules assigned to the first, second and third layers are shown in different colors in the space-filling model. We present the structures as obtained 1) through direct full atomic relaxation of the as-built trilayer and 2) through consecutive molecular dynamics (T was increased up to 400 K and then the system was cooled down again) and full atomic relaxation. We notice that, after MD, in model I, a molecule of the first layer enters into the z-value range of the second water layer, that the second layer becomes more disordered and less compact, whereas the third layer is largely reorganized, as shown in Figure 9a and c. In model II (Figure 9b and d), the first layer is rather stable. In the second, only one molecule is lifted up in the z-range of the third layer, and the third layer is kept tight to the second with no OH pointing towards the vacuum.

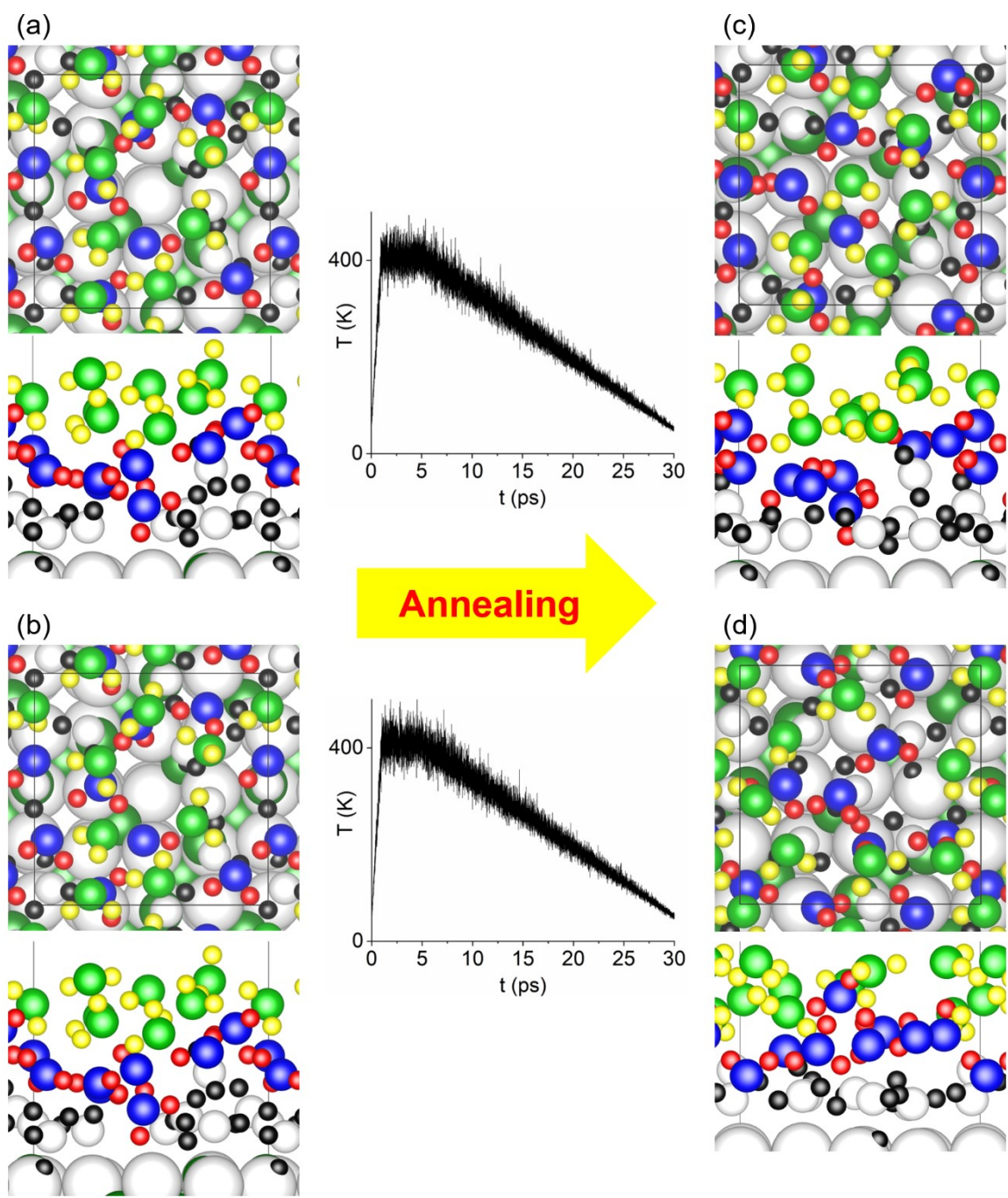

**Figure 9**. Top and side views of water trilayer adsorbed on the top $Fe_3O_4$(001) surface in two different configurations. (a) and (c) are model I before and after annealing, respectively. (b) and (d) are model II before and after annealing, respectively. The big green and big white balls represent Fe and O in the surface. The small white, blue and small green balls represent O in the first, second and third water layers, respectively. The black, red and yellow balls represent H in the first, second and third water layers, respectively. The black squares represent the ($\sqrt{2} \times \sqrt{2}$)R45° unit cell used in the calculations. Simulated annealing temperature profiles are inserted in the middle.

Similarly to what we did for the bilayer above, we have analyzed the water molecules distance and orientation from the surface by plotting the distribution functions 1) of the z coordinate of the water O atoms (Figure 10a and c) and 2) of the angle between the OH bonds direction and the surface normal (Figure 10b and d), for all the structures obtained at every step of the production run in the DFTB+U molecular dynamics simulation at 300 K for 50 ps. In Figure 10a and c, on the left, we present the distribution of the z coordinates of the O atoms and, on the right, we present their time dependence for all the twenty-four water molecules. In line with what found for the bilayer models, the third layer is less compact than the second that is less compact than the first (average z-value for O atoms in the third layer of model I [6.64±0.88 Å] and of model II [6.01±0.70 Å] vs that in the second layer of model I [4.35±0.61 Å] and of model II [3.81±0.57 Å] vs that in the first layer of model I [2.29±0.44 Å] and of model II [2.06±0.15 Å], respectively. Different layers here are defined according to the peaks in the distribution of the z coordinates of the O atoms as shown in Figure 10a and c.) Especially for model I, the broadening of the distribution peaks makes more difficult to define which molecules belong to the second and which to the third layer. This is a clear indication that as soon as we get further from the magnetite surface, we observe a loss of the ordering effect. In Figure 10b and d, we observe the orientation of the molecules in the three layers (grey, blue and green dots) for the two models proposed. In the first layer of both models, the water molecules are oriented in the plane of the layer but also downwards toward the surface atoms and upwards toward the second layer. This is an indication that they are forming all types of H-bonds: intralayer, interlayer and with the surface O atoms. The behavior of the second layer in both trilayer models is rather different from what observed in the bilayer models, where the water molecules were mostly lying in the layer plane. Here, on the contrary, their behavior is more similar to the first layer and the water molecules are oriented in all the directions because they form both intralayer and interlayer H-bonds again. The angle distribution for the third layer is less peaked at the center than what observed for the second layer in the bilayer models, which means that less molecules are lying in the layer plane because of a higher number of interlayer H-bonds. However, there is one OH from one water molecule in the third layer of model I that always points upwards during the MD simulation, which results in a peak at $\cos\theta = 1$ in the angle distribution plotting for the third water layer (Figure 10b).

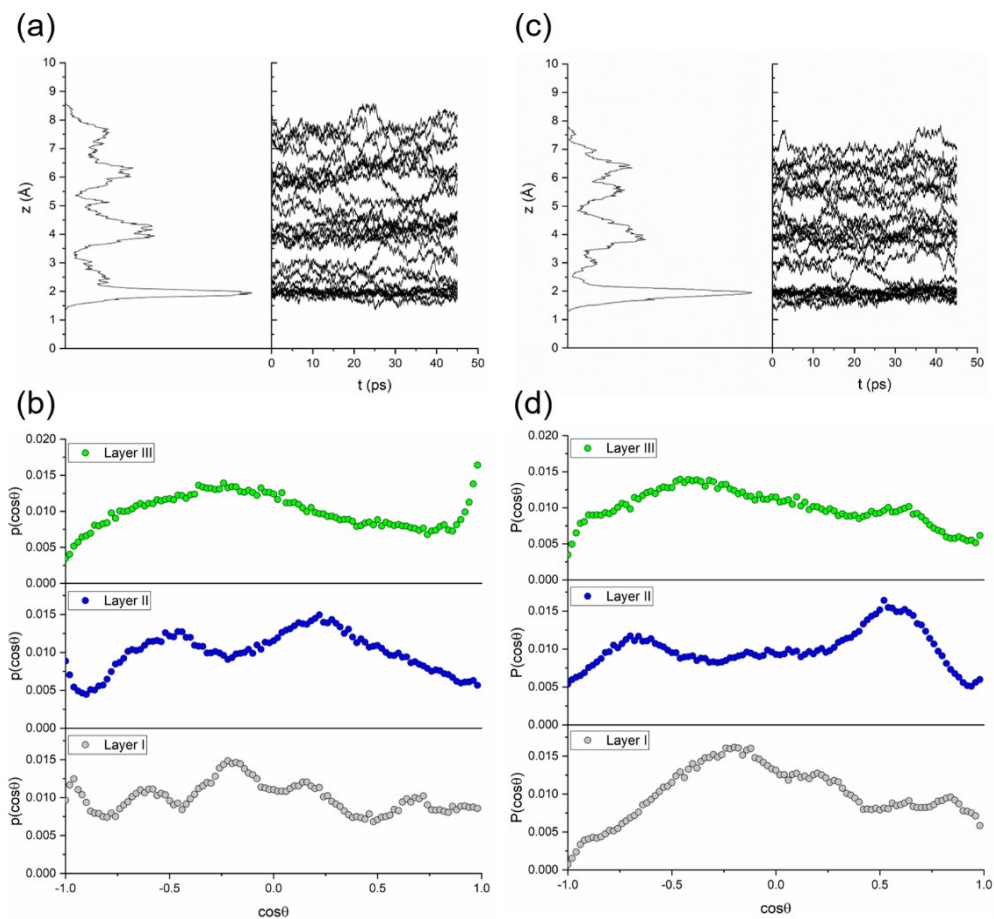

**Figure 10**. (a) and (c) are the distribution and time evolution of z (perpendicular distance from the surface) for the water oxygens of an adsorbed water trilayer on the $Fe_3O_4(001)$ surface for model I and model II, respectively. z is referred to the average value of z of the four $Fe_{Oct-5c}$ atoms on the surface. (b) and (d) are probability distribution P(cosθ) of the angle θ between the O-H bond vector and the normal of the surface for the molecules of the water trilayer adsorbed on the $Fe_3O_4(001)$ surface for model I and model II, respectively. Cosθ equals to 1 means that the O-H bond is directed upwards, whereas a value close to -1 means that the O-H bond is directed downwards.

The adsorption energy per water molecule ($E_{ads}$) is -0.68 and -0.70 eV for model I and II of trilayer structures, respectively. Therefore, we observe an inversion of stability: for the water monolayer, model I was the most stable; for the water bilayer, model I and II were isoenergetic; for the water trilayer system, model II becomes the most favorable. The adhesion energy per molecule ($E_{adh}$) of the third layer is larger for model II than for model I (-0.55 vs -0.49 eV, respectively). These adhesion energy values per molecule are lower than those computed for the first layer in the monolayer structure because the latter is directly bound to the magnetite surface, but are similar to (only slightly lower than) those computed for the second layer in the bilayer, where only H-bonds among water molecules were established, similarly to the present case.

There is a simple reason why the two models are not isoenergetic: the two trilayer structures present a different total number of H-bonds. For both models, we observe the formation of eight H-bonds in the first water layer and seven in the second layer. In the third layer, eleven and ten H-bonds are established in model I and II, respectively. Between the first two layers, seven and nine H-bonds are formed, whereas, between the second and third layers, we count four and seven, respectively. Finally, there are three and two H-bonds between the first layer and the surface, respectively. The sum of all these H-bonds is forty for model I and forty-three for model II of water trilayer.

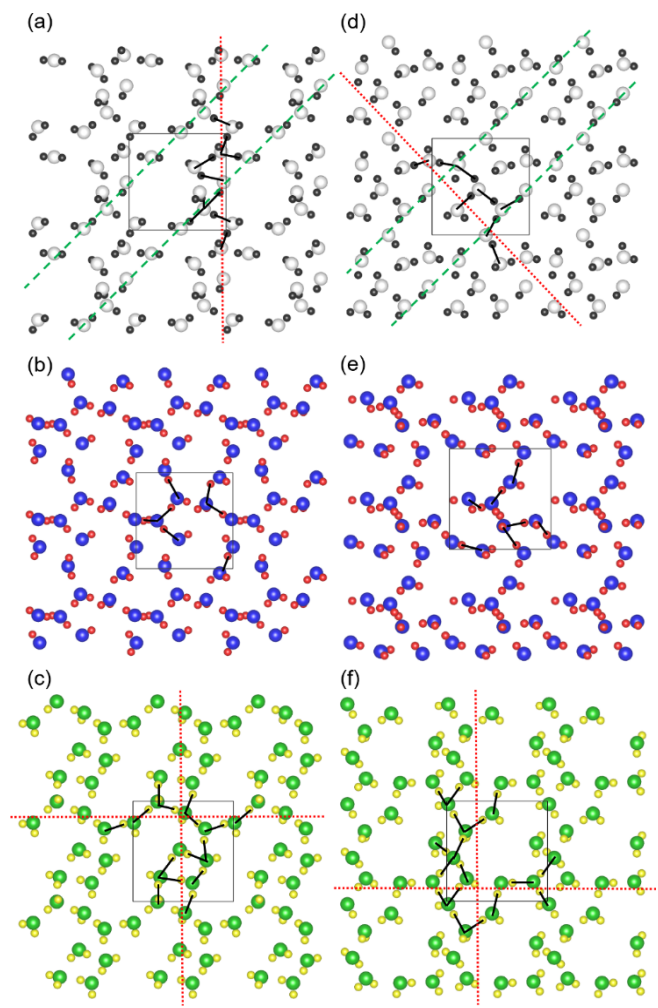

**Figure 11**. H-bond network in the first ((a) and (d)), second ((b) and (e)) and third ((c) and (f)) water layer adsorbed on the $Fe_3O_4$(001) surface for model I (left) and model II (right). The black segments represent the H-bonds as obtained at DFTB+U level of theory. The dotted red lines indicate the directions along which the H-bond networks develop. The dashed green lines indicate the [110] directions of $Fe_{oct-5c}$ atoms on the surface where the water molecules are adsorbed. The black squares represent the ($\sqrt{2} \times \sqrt{2}$)R45° unit cell used in the calculations.

We have analyzed whether there is some H-bond network existing in these two model structures, as presented in Figure 11. We investigated each layer, separately. For both models, in the first and third layers, we clearly observe in-plane H-bond networks that become infinite through the repetition of periodic supercells. On the contrary, in the second layer, no bidimensional H-bond network is established because the molecules are involved in H-bonds with water molecules of the first and the third layer. Therefore, the role of the second layer is to provide adhesion between the first and the third one, for instance, as an interconnecting layer.

### *3.4 Classical molecular dynamics simulations of a $Fe_3O_4$ (001) interface with a water trilayer and with bulk water*

When the number of layers further increases, the system becomes too large to be investigated at a quantum mechanical level of theory. For this reason we made recourse to the molecular mechanics (MM) approach, based on force fields, that allows for a much larger number of atoms (a thick water multilayer or even bulk water) and for a much longer simulation time length (in the range of ns). We filled the empty space between repeated images of the hydroxylated (4×4) supercell 17-layer slab along the z-direction with a ~12 nm-thick water multilayer (Figure 12a). Then, an MM-MD simulation was carried out for 10 ns at 300 K. We also reinvestigated the water trilayer on the $Fe_3O_4$ (001) surface for a comparative analysis with DFTB+U results, using the same MD simulation time length of 50 ps at 300 K.

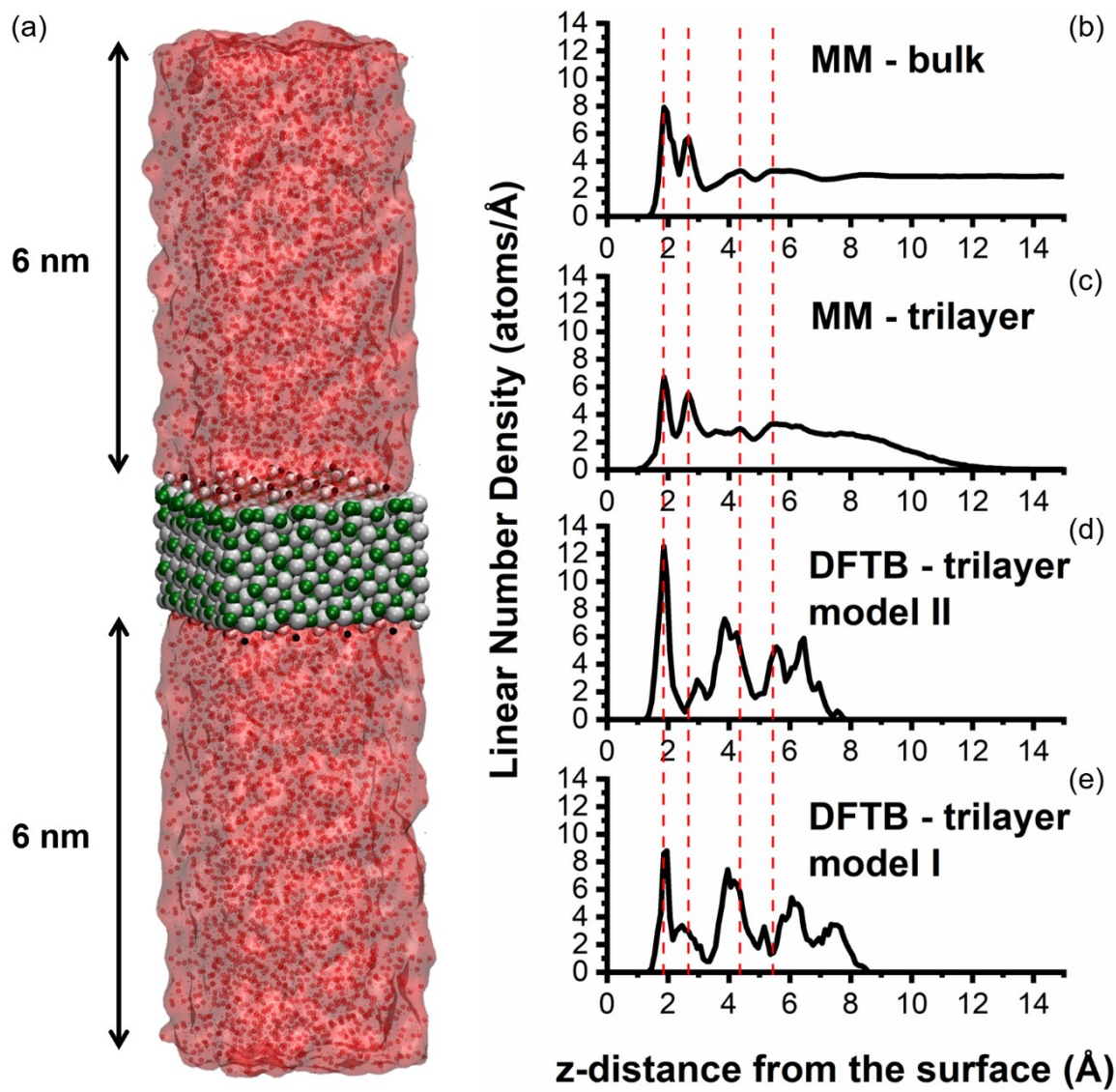

**Figure 12.** (a) Model of the hydroxylated (4×4) $Fe_3O_4$(001) slab interfaced with a ~12 nm-thick water multilayer in a periodic boundary condition fashion. Linear number density of water molecules, as a function of the distance from the slab surface, for the model interfaced with (b) a ~12 nm-thick water multilayer and with (c) a water trilayer, as averaged along all the MM-MD simulation run, and for (d) trilayer model II and (e) trilayer model I, as averaged along all the DFTB-MD simulation runs, respectively.

The results are shown in Figure 12. On the right side, we compare the linear number density of the water molecules at increasing distance (z-coordinate) along the surface normal with respect to the average z-coordinated of the surface $Fe_{Oct}$ ions, as averaged along all the MM-MD simulation run, for the slab interfaced with ~12-nm water multilayer (Figure 12b) and for the slab interfaced with a water trilayer (Figure 12c) with the corresponding distributions obtained with DFTB-MD simulation for the trilayer model II (Figure 12d) and trilayer model I (Figure 12e) from the previous sections.

The red dashed lines are aligned to the center of the four peaks in the distribution for the slab model in bulk water (at 1.9, 2.7, 4.4 and 5.5 Å, Figure 12b). This calculation can be considered as the reference since the size of the model and the simulation time length are the largest of all. The distribution of the water density converges correctly to the value of bulk water (0.33 e/Å$^3$), as shown in Figure S3. We observe that the peaks for the MM water multilayer on the $Fe_3O_4$ (001) surface slab are centered at the same positions as for the MM water trilayer (Figure 12b vs 12c). Besides the fact that for the MM water trilayer we observe an additional small feature at about 3 Å, the two plots are very similar, especially as regards the first two features: they are centered at the same position, the height of the peaks is similar, especially for that at 2.7 Å (see also Figure S4 for a direct comparison). These two features refer to the first layer of water molecules in direct contact with the surface. The first peak centered at 1.9 Å is due to the water molecules that are H-bonded (as acceptors) to the surface OH or (as donors) to surface O atoms. The second peak centered at 2.7 Å is due to the water molecules that are directly coordinated to a surface $Fe_{Oct}$ atom ($Fe_{Oct}$---$OH_2$). Therefore, we may conclude that, although the water trilayer model is clearly a limited one, it may capture the main features of the interfacing water in direct contact with the surface providing results in excellent agreement with a model of bulk water on the surface.

We now compare (in Figure 12c vs 12d and 12e) the results for the water trilayer on the $Fe_3O_4$ (001) surface from MM with those from DFTB+U calculations that have been discussed in the previous sections (see also Figure 10). Since DFTB+U method has been validated on the base of a comparison with DFT results in Section 3.1, we consider them as a good benchmark reference for MM calculations. There is a first clear difference between the MM-MD trilayer curve (Figure 12c) and the DFTB+U ones (Figures 12d and 12e) that is related to the fact that the first two peaks observed for the MM-MD calculation merge into one single peak just below 2 Å for the DFTB+U-MD (model II) simulation (Figure 12d). We have proved this by analyzing the types of water molecules that make up the first high peak in the DFTB+U-MD (model II): both water molecules that are H-bonded to the surface and that coordinated to the surface Fe ions are included in this first feature because the average distance in both cases is observed to be at about 1.9 Å. In the case of the DFTB+U-MD (model I) simulation (Figure 12e), we observe an additional low broad feature at a distance from the surface between 2 and 3 Å, which is assigned to some H-bonded water molecules that lie between the first layer of water (directly interacting with the surface magnetite atoms) and the second layer of water. In the case of model II, these straddling water molecules are observed at about 3 Å.

We can provide a clear explanation of the reason why the peak is split in two features in the MM-MD simulation curve. This is because of the poor transferability of the cross-term LJ 9-6 parameters (calculated by the sixth-power combining rule) of the Fe(III) atom[35] with the COMPASS-based three-site water model[36] for the description of the Fe---$OH_2$ cross interaction, which leads to a longer and weaker coordinative bond compared with QM data. One can check a set of tests of the MM model in the supplementary material. We are working on the parametrization of this model, and the results will be the subject of a future study.

A second difference between the MM-MD trilayer curve and the DFTB+U-MD ones is the broader range of z-values that are reached during the MM-MD simulation that indicates a higher tendency of water molecules to diffuse towards the vacuum.

## 4. Conclusions

In conclusion, through this multiscale computational study on the $Fe_3O_4$ (001) surface/water interface, we have first assessed the reliability of the DFTB to accurately describe the structural details and the energetics as compared to hybrid functional HSE06 calculations, then we have investigated different models of water bi- and trilayers, through DFTB-MD simulations, and, finally, we have compared these results with long time scale and large size MM simulations for a thick (12 nm) water multilayer sandwiched between two $Fe_3O_4$ (001) slabs. The most stable configuration for an adsorbed water monolayer (model I), that was previously confirmed by experimental STM results, here is not found to be the most favorable molecular assembling to build up water bi- and tri-layers. Water molecule become more packed when additional overlayer are added. The water structuring into molecular layers observed with DFTB-MD simulations for the proposed bi- and tri-layers models is confirmed by longer MM-MD simulations on larger and more realistic systems of $Fe_3O_4$ (001) surface/ bulk water interface. However, the combination of the existing MM parameters leads to the overestimation of the Fe---$OH_2$ distance from the surface and, thus, calls for an improvement. The present multiscale study is a good starting point to pursue this goal.

AUTHOR INFORMATION

Corresponding Author

*E-mail: cristiana.divalentin@unimib.it

**SUPPLEMENTARY MATERIAL**

See supplementary material for further computational details, tables with comparative analysis, figures of the structures for three and six water molecules adsorbed on the top $Fe_3O_4(001)$ surface, tables with additional structural information, electron density profile of bulk water on the $Fe_3O_4(001)$ surface (against liquid) and comparative linear number density profiles of all the systems under investigation.

**ACKNOWLEDGMENTS**

The authors are grateful to Lorenzo Ferraro for his technical help. The project has received funding from the European Research Council (ERC) under the European Union's HORIZON2020 research and innovation programme (ERC Grant Agreement No [647020]).

Supplementary Material

# Insight into the interface between $Fe_3O_4$ (001) surface and water overlayers through multiscale molecular dynamics simulations

Hongsheng Liu[1,2], Enrico Bianchetti[1], Paulo Siani[1], Cristiana Di Valentin[1*]

1. Dipartimento di Scienza dei Materiali, Università di Milano-Bicocca

via Cozzi 55, 20125 Milano Italy.

2. Key Laboratory of Materials Modification by Laser, Ion and Electron Beams,

Dalian University of Technology, Ministry of Education, Dalian 116024, China

[*] Corresponding author: cristiana.divalentin@unimib.it

**Further computational details:**

*Testing the COMPASS-based Fe$_3$O$_4$(001)/water model*

To test the transferability of the LJ(9-6) COMPASS-FF parameters[1] for Fe(II), Fe(III) and O(II) atoms with the COMPASS-based three-site water model and hydroxyl group parameters available in the INTERFACE-FF,[2] we rely on a direct comparison of adsorption energies and structural parameters of the MM predictions against the QM data. One should keep in mind that only energy minimization of the molecular water has been considered in these calculations.

It is also worth recalling the several approximations imposed on our MM model:

1) The partial atomic charges for the Fe(III), Fe(II) and dissociated water atoms were derived at the DFT/HSE06 level of theory, and the partial charges for the O(II) atoms were calculated by neutralizing the remaining positive charge;

2) Slab atoms in the MM model were frozen at their optimized position obtained at DFTB+U level of theory, and only non-bonded interactions between the Fe$_3$O$_4$(001)/Fe$_3$O$_4$(001) and Fe$_3$O$_4$(001)/water atoms were considered;

3) All pairwise LJ-interactions in the system followed the sixth-power combining rule,[3] in line with the CLASS2-FFs philosophy.

Herein, we have tested both the non-hydroxylated and the hydroxylated surface. The former model is used to represent the physisorption of one water molecule on the bare surface. The latter model is used to represent the physisorption of one water molecule in the presence of one dissociated water molecule, leading to the formation of a stable water dimer species on the surface. Since the MM model cannot describe the dissociation process, we took the atomic positions for the hydroxylated surface from DFTB+U and kept them fixed.

The adsorption energies ($E_{ads}$) were estimated by subtracting the total potential energy of the isolated Fe$_3$O$_4$(001) slab model ($E_{slab}$), either hydroxylated or non-hydroxylated, and that of one isolated water molecules ($E_{H_2O}$) from the total potential energy of the adsorbed surface ($E_{total}$):

$$E_{ads} = E_{total} - E_{slab} - E_{H_2O}$$

Table S1 lists the energetic and structural data obtained at MM and QM level of theory. The numbers in parenthesis stand for the percentage error of adsorption energies and inter-atomic distances calculated at MM level against DFT data.

**Table S1.** Adsorption energies of one water molecule on non-hydroxylated and hydroxylated $Fe_3O_4$ (001) surfaces.

| Adsorption energy (eV) | DFTB+U (DFTB+) | DFT/HSE06 (CRYSTAL17) | MM/CLASS2-FF (LAMMPS) |
|---|---|---|---|
| 1 $H_2O$ (Molecular) | -1.14 | -0.94 | -0.66 (29.8%) |
| 2 $H_2O$ (Half dissociated) | -1.54 | -1.50 | -1.28 (14.7%) |

**Table S2.** Structural parameters of one water adsorbed on non-hydroxylated and hydroxylated $Fe_3O_4$ (001) surfaces.

| Inter-atomic distance (Å) | | DFTB+U (DFTB+) | DFT/HSE06 (CRYSTAL17) | MM/CLASS2 (LAMMPS) |
|---|---|---|---|---|
| 1 $H_2O$ (Molecular) | Fe - $OH_2$ | 2.23 | 2.16 | 2.84 (31.5%) |
| 2 $H_2O$ (Half dissociated) | Fe - OH | 1.89 | 1.93 | 1.89* |
| | Fe - $OH_2$ | 2.01 | 2.06 | 2.69 (30.6%) |
| | HO - - H – OH | 1.57 | 1.48 | 1.62 (9.5%) |
| | HO - - $OH_2$ | 2.57 | 2.52 | 2.60 (3.2%) |

*Fixed atoms in the slab.

Table S1 shows the adsorption energy of a single water molecule on both non-hydroxylated and hydroxylated $Fe_3O_4$ (001) surfaces at MM and QM levels of theory. The adsorption energy of a single molecule on the non-hydroxylated surface has a significant deviation from the reference data. On the other hand, for the hydroxylated $Fe_3O_4$ (001) surface, the adsorption energy of a single water molecule shows a deviation ~15% from the reference data. This fact could be attributed to the presence of the hydroxyl groups on the hydroxylated $Fe_3O_4$ (001) surface that stabilizes the water molecule via H-bond.

H-bond distances between the hydroxyl O atom and the H atom of molecular water are in good agreement with the QM data. Furthermore, the distance between the hydroxyl O atom and the O atom of water as well as the angle of hydrogen bond formed (DFTB: 162.12°, DFT: 169.94°; MM:

165.57°) are both in line with the QM predictions. On the other hand, the present MM model predicts a longer inter-atomic distance for the Fe(III)-OH$_2$ pair than that obtained at the QM level of theory. This discrepancy may arise in part from the aforementioned approximations imposed on theMM model. For instance, there is the loss of degrees of freedom caused by keeping the Fe$_3$O$_4$ (001) slab atoms frozen and there is a lack of atomic polarizability, which deserves especial attention for further improvement of the MM model.

This set of tests provides a deeper understanding of the COMPASS-based Fe$_3$O$_4$(001)/water model as well as a reliable starting point for further parameterization and validation. We are currently working along this line on the parametrization of the Fe$_3$O$_4$(001)/water interface model and the results will be the subject of a future work.

*Density profile calculations*

For the sake of comparison, linear number density profiles (atoms/Å) were calculated to compare both the MM-MD and DFTB-MD results, in which only O atoms belonging to the molecular water were considered. First, we have divided the space along the z coordinate in equally-sized bins (Δz) of thickness set at 0.1 Å. Then, the bins count was divided by the total count (number density) and normalized by the bin size. Figure 12 in the manuscript shows the linear number density profiles calculated from the DFTB-MD and MM-MD trajectories. Furthermore, we have also normalized the density profile of the MM-bulk model to the electron density unit (e/Å$^3$) to make its convergence to the experimental value of liquid water clear, as shown in Figure S3 below.

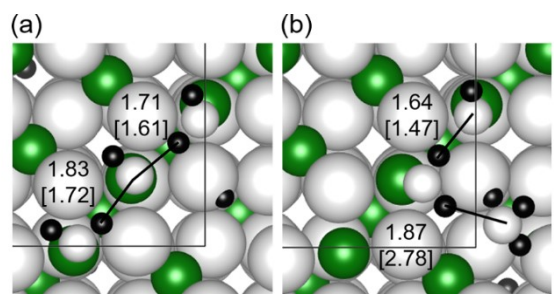

**Figure S1.** Top views of Fe$_3$O$_4$(001) surface with three water molecules adsorbed on the top in different configurations: (a) "linear" adsorption mode and (b) "nonlinear" adsorption mode. The green, black, small white and big white balls represent Fe, H, O from water and O from the surface. The hydrogen bonds are indicated by the solid lines. The bond length calculated by DFTB+U (without any brackets) and DFT/optPBE-DF+U (inside the square brackets) are given for each configuration. The black squares represent the ($\sqrt{2} \times \sqrt{2}$)R45° unit cell used in the calculations.

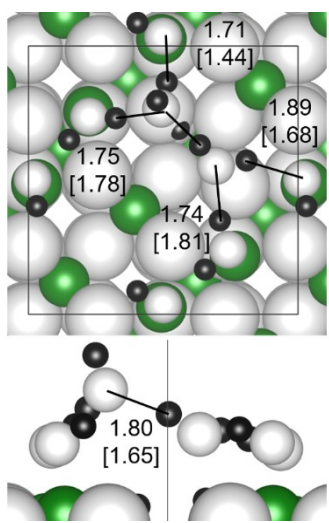

**Figure S2.** Top and side view of Fe$_3$O$_4$(001) surface with six water molecules adsorbed on the top. The green, black, small white and big white balls represent Fe, H, O from water and O from the surface. The hydrogen bonds are indicated by the solid lines. The bond length calculated by DFTB+U (without any brackets) and DFT/optPBE-DF+U (inside the square brackets) are given for each configuration. The black square represent the ($\sqrt{2} \times \sqrt{2}$)R45° unit cell used in the calculations.

**Table S3.** The bond lengths of Fe-O$_w$ (O$_w$ represents the O atom in water) in water/Fe$_3$O$_4$(001) systems obtained with different methods. The bond length obtained by DFT/HSE06 method is taken as the reference for calculating the error. The average error for all the configurations is 2.4%.

| Coverage | Configurations | Bond length (Å) | | Error (%) |
|---|---|---|---|---|
| | | DFTB+U | DFT/HSE06 | |
| 1 H$_2$O | Molecular | 2.23 | 2.16 | 3.2 |
| | Dissociated | 1.79 | 1.82 | 1.6 |
| 2 H$_2$O | Mixed | 1.89 | 1.93 | 2.1 |
| | | 2.01 | 2.06 | 2.4 |
| | Molecular | 2.26 | 2.24 | 0.9 |
| | | 2.25 | 2.18 | 3.2 |
| | Dissociated | 1.80 | 1.83 | 1.6 |
| | | 1.80 | 1.83 | 1.6 |
| 4 H$_2$O | Mixed | 2.24 | 2.09 | 7.2 |
| | | 1.89 | 1.94 | 2.6 |
| | | 2.08 | 2.09 | 0.5 |
| | | 1.90 | 1.94 | 2.1 |
| | Molecular | 2.27 | 2.22 | 2.3 |
| | | 2.27 | 2.23 | 1.8 |
| | | 2.27 | 2.31 | 1.7 |
| | | 2.27 | 2.31 | 1.7 |
| | Dissociated | 1.83 | 1.85 | 1.1 |
| | | 1.82 | 1.84 | 1.1 |
| | | 1.82 | 1.83 | 0.5 |
| | | 1.81 | 1.83 | 1.1 |
| 8 H$_2$O | Model I | 1.91 | 1.98 | 3.5 |
| | | 1.89 | 1.94 | 2.6 |
| | | 2.24 | 2.12 | 5.7 |
| | | 2.07 | 2.08 | 0.5 |
| | Model II | 1.90 | 1.96 | 3.1 |

| | | 1.89 | 1.94 | 2.6 |
| | | 2.24 | 2.10 | 6.7 |
| | | 2.06 | 2.08 | 1.1 |

**Table S4.** The H-bond lengths in water/$Fe_3O_4$(001) systems obtained with different methods. The bond length obtained by DFT/HSE06 method is taken as the reference for calculating the error. The average error for all the configurations is 8.2%.

| Coverage | Configurations | Bond length (Å) | | Error (%) |
|---|---|---|---|---|
| | | DFTB+U | DFT/HSE06 | |
| 2 $H_2O$ | Mixed | 1.57 | 1.48 | 6.1 |
| | Molecular | 2.06 | 2.24 | 8.0 |
| | Dissociated | - | - | - |
| 4 $H_2O$ | Mixed | 1.70 | 1.51 | 12.6 |
| | | 1.65 | 1.51 | 9.3 |
| | Molecular | 2.24 | 2.37 | 5.5 |
| | | 2.07 | 2.01 | 3.0 |
| | | 2.29 | 1.94 | 18.0 |
| | | 2.02 | 2.65 | 23.8 |
| | Dissociated | 2.36 | 2.16 | 9.3 |
| | | 2.07 | 2.15 | 3.7 |
| | | 2.46 | 2.26 | 8.8 |
| | | 2.06 | 2.13 | 3.3 |
| 8 $H_2O$ | Model I | 1.81 | 1.88 | 3.7 |
| | | 1.72 | 1.54 | 11.7 |
| | | 1.81 | 1.68 | 7.7 |
| | | 1.79 | 1.76 | 1.7 |
| | | 1.80 | 1.72 | 4.7 |
| | | 1.76 | 1.58 | 11.4 |
| | | 1.73 | 1.62 | 6.8 |
| | | 1.76 | 1.77 | 0.6 |
| | | 1.74 | 1.60 | 8.8 |
| | | 2.13 | 2.32 | 8.2 |
| | | 2.11 | 2.00 | 5.5 |
| | | 1.75 | 1.79 | 2.2 |
| | | 1.77 | 1.68 | 5.4 |
| | | 2.14 | 1.80 | 18.9 |
| | Model II | 1.86 | 1.88 | 1.1 |
| | | 1.79 | 1.59 | 12.6 |
| | | 1.94 | 2.43 | 20.2 |
| | | 1.82 | 1.80 | 1.1 |
| | | 2.10 | 1.78 | 18.0 |
| | | 1.76 | 1.62 | 8.6 |
| | | 1.73 | 1.59 | 8.8 |
| | | 2.10 | 1.96 | 7.1 |
| | | 1.72 | 1.92 | 10.4 |
| | | 2.18 | 2.02 | 7.9 |
| | | 1.71 | 1.65 | 3.6 |
| | | 1.72 | 1.66 | 4.2 |

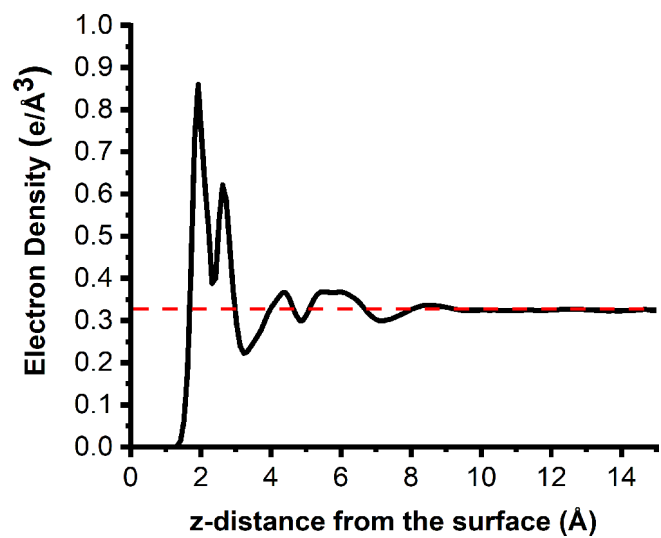

**Figure S3.** Electron density profile of water molecules, as a function of the distance from the slab surface, for the model interfaced with ~12-nm water multilayer. The red dashed line represents the experimental value for liquid water.[4]

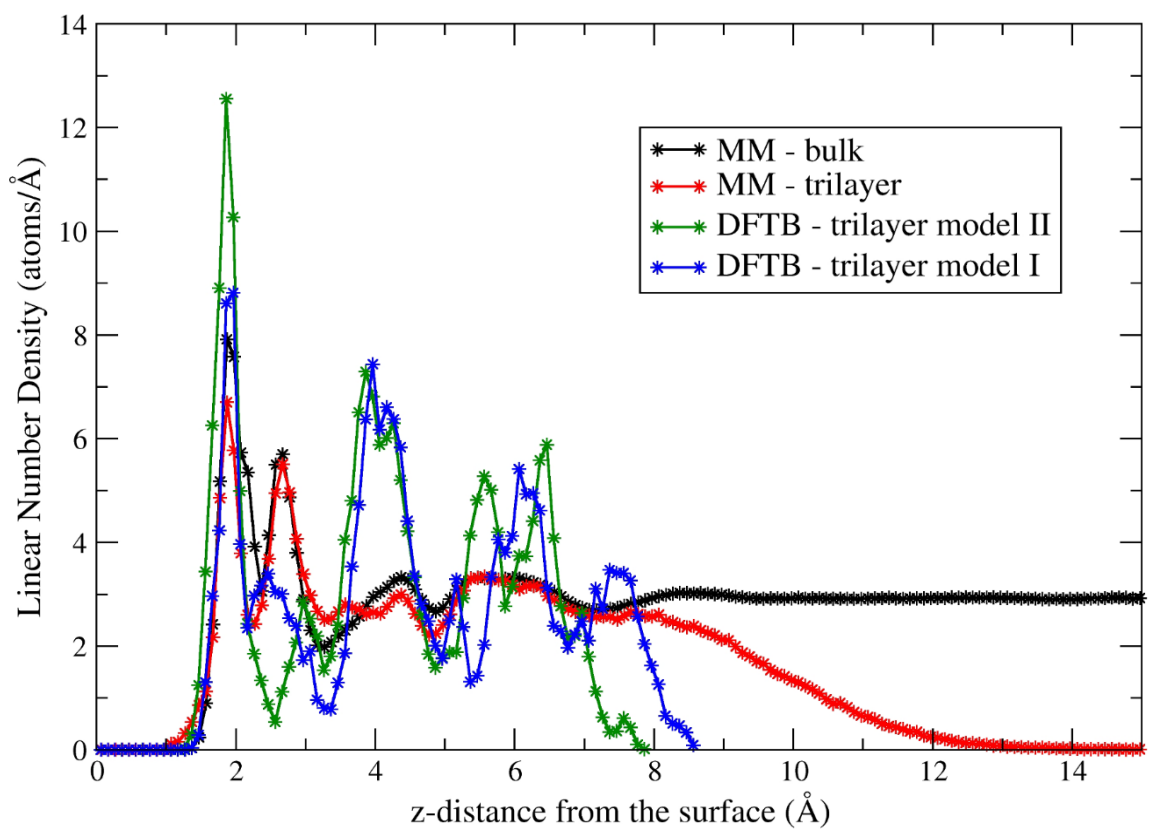

**Figure S4.** Linear number density of water molecules, as a function of the distance from the slab surface, for the model interfaced with a ~12 nm-thick water multilayer and with a water trilayer, as averaged along all the MM-MD simulation run, and for trilayer (model II) and trilayer (model I), as averaged along all the DFTB-MD simulation run, respectively.